\RequirePackage{fix-cm}
\documentclass[twocolumn]{svjour3}          
\smartqed  
\usepackage{graphicx}
\usepackage{mathptmx}      
\begin{document}

\title{Laser ablation in liquid: bridge from a plasma stage to bubble formation
}


\author{
N.~A. Inogamov \and V.~A. Khokhlov \and Yu.~V.~Petrov \and V. V. Zhakhovsky }


\institute{
N. A. Inogamov, \email{nailinogamov@gmail.com}, V.A. Khokhlov, Yu. V. Petrov \at Landau Institute for
Theoretical Physics of the Russian Academy of Sciences, Akademika Semenova 1a, Chernogolovka, Moscow Region
142432, Russia \and V. V. Zhakhovsky, N. A. Inogamov \at Dukhov Research Institute of Automatics (VNIIA),
Sushchevskaya 22, Moscow 127055, Russia \and V. V. Zhakhovsky \at Joint Institute for High Temperatures of
Russian Academy of Sciences, Izhorskaya st. 13 Bd.2, Moscow 125412, Russia \and Yu. V. Petrov \at Moscow
Institute of Physics and Technology, Institutskiy Pereulok 9, Dolgoprudny, Moscow Region 141701, Russia }


\maketitle

\begin{abstract}
Laser ablation through liquid is an important process that have to be studied for applications which use laser
ablation in liquid (LAL) and laser shock peening (LSP). LAL is employed for production of suspensions of
nanoparticles, while LSP is applied to increase hardness and fatique/corrosion resistance properties of a
surface layer. A bubble appears in liquid around the laser spot focused at a target surface after strong enough
laser pulse. In the paper we connect the early quasi-plane heated layer created by a pulse in liquid and the
bubble forming at much later stages. In the previous works these early stage from one side and the late stage
from another side existed mainly as independent entities. At least, quantitative links between them were
unknown. We consider how the quasi-plane heated layer of liquid forms thank to thermal conduction, how
gradually conduction becomes weaker, and how the heated layer of liquid nearly adiabatically expands to few
orders of magnitude in volume during the drop of pressure. Our molecular dynamics simulations show that the
heated layer is filled by the diffusive atomic metal-liquid mixture. Metal atoms began to condense into
nanoparticles (NP) when they meet cold liquid outside the edge of a mixing zone. This process limits diffusive
expansion of metal atoms, because diffusive ability of NP is less than this ability for individual atoms. Thus
the mixture expands together with hot liquid, and the NPs approximately homogeneously fill an interior of a
bubble.
\end{abstract}

\section{Introduction}
\label{intro}
 Laser ablation in liquid (LAL) is important for many technological applications listed in the recent reviews
  \cite{LZ+Stephan:2018.LAL,XIAO:2017} and papers
 \cite{nano9101348,Krylach_2019,doi:10.1063/1.5114630,doi:10.1021/acsnano.9b04610,Bally:2019,C8AN01276H}.
 Another important direction of industrial applications is connected with laser shock peening (LSP)
\cite{obzor:LSP:2017,Veiko:2017,Kudryashov:2019:UFast,Kudryashov:2019:Filam,CORREA:2015}.
 Theoretical analysis of LSP is related to the analysis of laser initiated shock waves.
 There are papers about elastic-plastic penetration of shocks
\cite{Ashitkov:jetpLett:2010,superElast:2010okt,Basil:1W2Z:PRL:2011,Kanel:elast-plast-SW:2012,Kanel:1426,Demaske:PRB:2013,Kanel:Ag:2014}.
 Nanosecond LSP is performed in water environment - water flows along surface of a target.
 This water surrounding is dynamically important feature of LSP.
 This feature increases amplitude of a shock
  \cite{obzor:LSP:2017,CORREA:2015,Fabbro:1990}
   and momentum transferred to a target.
 Thus LSP is similar to laser ablation in liquid.  


 LSP and LAL are two sides of the same process of laser interaction with target through transparent liquid.
 In LSP the people are interested in processes going inside a target - evolution of ejecta and production of nanoparticles (NP) are out of their area of interest.
 In femtosecond (fs) LAL \cite{T.Sano:2017,T.Sano:2018} the situation is opposite - shock runs away fast from a heat affected zone (tens of fs).
 While evolution of ejecta (laser plume) continues for a long time
  \cite{LZ+Stephan:2018.LAL,Povarnitsyn-Itina:LAL:2013,povarnitsyn:ITINA:LAL:2014,LZ-bulk-LAL:2017,INA.jetp:2018.LAL,Petrov:LAL:Appl.Surf.Sci:2019,Petrov:LAL:Contrib.Plasma.Phys:2019}.
 Namely this evolution is responsible for production of colloid of nanoparticles (NPs).



 Amplitude of compressive shock is independent on presence or absence of liquid in fs LSP (if absorbed fluences are equal);
   this is true namely for fs LAL; but for ns LAL the presence of liquid affects compressive shock inside a target amplifying it.
 In nanosecond (ns) LAL \cite{Petrov:LAL:Appl.Surf.Sci:2019,Petrov:LAL:Contrib.Plasma.Phys:2019} the interaction between shocks in liquid and in target
   continues during a pulse (ns time scale).
 But again after finishing of a ns pulse the shocks in target and in liquid go far away from the metal/liquid contact zone.
 While motions which produce colloids slow evolve around this contact zone up to ~1 microsecond \cite{Petrov:LAL:Appl.Surf.Sci:2019,Petrov:LAL:Contrib.Plasma.Phys:2019}.


 In experiments (devoted to LAL) using high-speed photo-imaging, the formation, expansion, stopping of expansion, and the collapse of a bubble in a liquid around a laser spot are observed
  \cite{Amans:2016:APL,LAL:exprmnt:bubble+spectr:2017,Amans:2019,Kanitz_2019}.  
 Appearance of the bubble is observed starting from $\sim 1$ microsecond $(\mu$s).
 Until times of the order of fractions of a $\mu$s, radiation emanating from the vicinity of the focusing spot is observed
   \cite{LAL:exprmnt:bubble+spectr:2017,Amans:2019,Kanitz_2019}. 
 The early stages of LAL of the order of ns remain experimentally uninvestigated.
 In other experiments, people follow development of optical breakdown of a laser beam while the beam approaches to surface of a target
   \cite{Bulgakov:2017,Gurevich:2019a,Gurevich:2019b}.


 There are relatively recent papers devoted to theory and numerical modeling of LAL
  \cite{Povarnitsyn-Itina:LAL:2013,povarnitsyn:ITINA:LAL:2014,LZ-bulk-LAL:2017,INA.jetp:2018.LAL,LZ+Stephan:2018.LAL,Petrov:LAL:Appl.Surf.Sci:2019,Petrov:LAL:Contrib.Plasma.Phys:2019}.
  In these works, the plasma stages of the development of processes are clearly traced.
  But these works are limited by short times to a few nanoseconds.
  The most important issue is the question of the relationship between the initial stages and the stage with the bubble.


 Here we describe the connection between the initial stage with the quasi-plane plasma flow on the one side
   and the final stage with the expanding bubble on the other side.
 The analysis given below in the paper is divided into three main parts.

 First, we present the data of calculating (by the hydrodynamic code 1D-2T-HD) the nanosecond laser action on gold through water
   up to the times at which the spherization of a shock wave front in water begins;
    spherization means transit from quasi-plane to quasi-spherical shape.

 Secondly, we construct families of water adiabats that cover our range of entropies and pressure drops during expansion.
 This range is from the order of $10^4$ bar to pressures of 0.1-1 bar.

 Thirdly, we present the data of molecular dynamics (MD) modeling regarding the mutual diffusion of gold and water
    together with the following picture of the condensation of gold atoms from MD.

\begin{figure}
  \centering \includegraphics[width=\columnwidth]{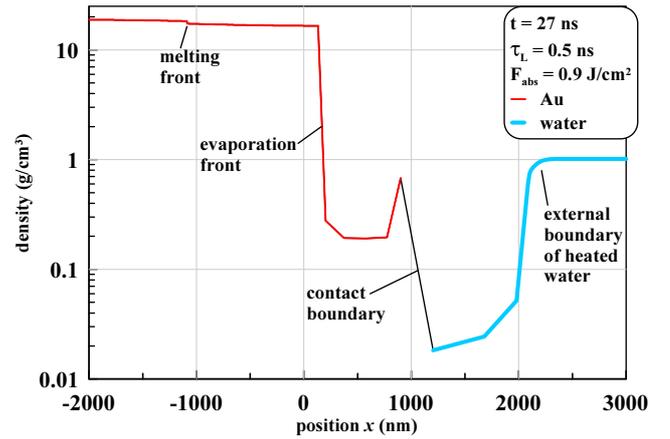}
\caption{Density profile at time 27 ns is shown. Duration $\tau_L$ and absorbed energy $F_{abs}$ of the laser pulse are given.
 Melting and evaporation fronts in gold are marked.
 The contact boundary separating gold and water is shown.
 We see the contact layers of evaporated gold and hot water.
 Water in the near-contact layer has expanded 30–40 times at a given time.
 The pressure at the indicated time 27 ns is still above the critical pressure for water, see next figure.
 Therefore, all water from the contact to the front of the shock wave is in a supercritical state.
 Thus, all water is in a homogeneous state, the opposite in this sense to gold.
 Position is reckoned from the initial (prior to laser action) contact between Au and water.
\label{fig:1}}       
\end{figure}

 Before describing these three parts of the work, we present data on the intermediate stage.
 At the intermediate stage, we will move from the description of the LAL flow using 1D-2T-HD code
   to the description using adiabatic curves.

\section{Intermediate stage}
\label{sec:2}



 The 1D-2T-HD code is a Lagrangian \cite{Samarskii:TheTheorDiffSc} 
 code in one dimension including full two-temperature (2T) physics
  \cite{Inogamov:2009:ASS,INA.jetp:2018.LAL,Shepelev:2019}. 
 This means that electrons are much hotter than ions. 
 For nanosecond (ns) laser pulse, the 2T effects are not significant \cite{Petrov:LAL:Appl.Surf.Sci:2019,Petrov:LAL:Contrib.Plasma.Phys:2019}.
 But code is universal in the sense that it may be used also for description of actions of ultrashort pulses where 2T physics plays dominate role.
 The code allows us to follow evolution of LAL up to few tens ns.
 To best of our knowledge, this is done for the first time in studies concerning LAL.

\begin{figure}
  \centering \includegraphics[width=\columnwidth]{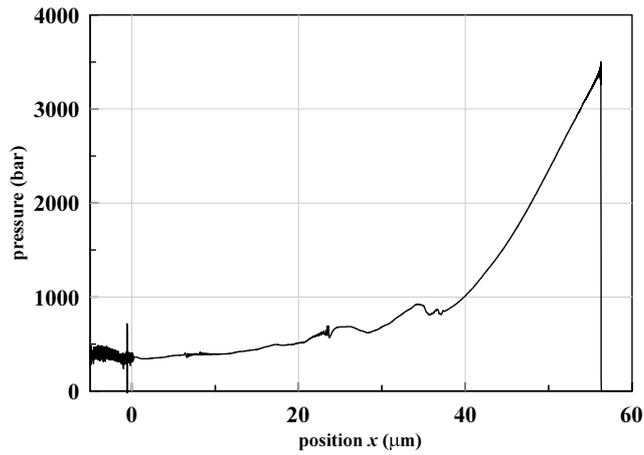}
\caption{The figure shows the entire area that is in motion at time 27 ns.
The parameters of the laser pulse are shown in Figure 1.
On the right edge is a shock wave in the water.
The near-contact layer is located near zero coordinates. The pressure gradients in the vicinity of the contact are small.
Contact pressure is much less than the pressure behind the shock wave.
The shock wave gradually attenuates and therefore acquires a triangular shape.
The pressure in the shock wave is much less than the bulk modulus 2.25 GPa for water.
 Therefore, the wave can be approximately described in the acoustic approximation. The speed of the wave is close to the speed of sound.
}
\label{fig:2}       
\end{figure}

\begin{figure}
  \centering \includegraphics[width=\columnwidth]{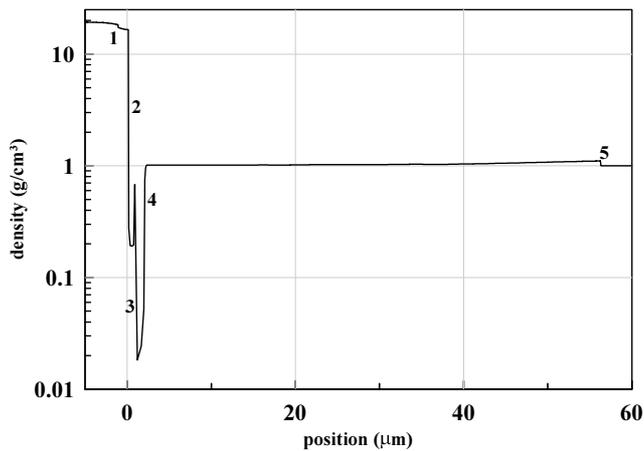}
\caption{The density profile at time 27 ns is shown here.
The parameters of the laser pulse are shown in Figure 1.
The contact layer 2-3-4 is tiny compared to the distance 5 the shock wave has traveled.
 Mark 1 denotes the instantaneous position of the melting front. The shock wave in gold went deep into the target.
 A rest zone remains behind it. Therefore, this figure shows the steam 2-3 and liquid 1-2 phases of gold,
   as well as part of the solid phase to the left of mark 1.
}
\label{fig:3}       
\end{figure}


 The main advantage in using the hydrodynamic code 1D-2T-HD is the use of adequate equations of state (EoS) substances.
 For gold the EoS is taken according to classical papers 
  \cite{Bushman:1993,Khishchenko2002,lomonosov_2007}, see also database \cite{rusbank1,rusbank2}.
 Therefore, phase transitions are described in a continuous manner inside the flow field.
 For problems concerning the LAL, descriptions of melting, recrystallization, and evaporation are of exceptional importance.
 The code clearly tracks melting and evaporation fronts.

\begin{figure}
  \centering \includegraphics[width=\columnwidth]{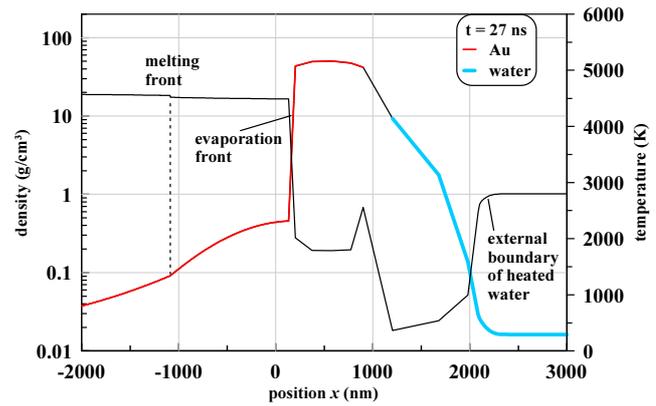}
\caption{Temperature distribution in the contact layers of gold and water.
A jump in the density and temperature gradient takes place at the melting front.
At the shown stage, the spreading of the melting front into gold is stopped.
That is, melting is replaced by re-crystallization.
The temperature jump at the evaporation front is associated with a strong drop in thermal conductivity in gaseous gold
 compared with the values of thermal conductivity in condensed gold.
The parameters of the laser pulse are shown in figure 1.
}
\label{fig:4}       
\end{figure}


A typical example is shown in the figure \ref{fig:1}.
 It presents vicinity of the gold-water contact.
 Namely this vicinity gives finite contribution into production of nanoparticles.
 There is a huge difference in the speed of sound in Au and water on the one side
   relative to the speed of expansion of the hot layer around the contact on the other side.
 Therefore, it is not surprising that the contact layer is much thinner than the distances traveled by the shock waves,
   see figures \ref{fig:2} and \ref{fig:3}.


 Figure \ref{fig:4} shows how hot the contact layers remain at times of the order of tens of nanoseconds.
 On the right outside the contact layer, the temperature quickly enough reaches the room temperature of the water.
 Since the dissipative heating of water in the shock wave is weak.
 Water heating occurs due to thermal conductivity.
 The transition layer from hot to cold water is indicated as "external boundary of heated water" in figures \ref{fig:1} and \ref{fig:4}.

\begin{figure}
  \centering \includegraphics[width=\columnwidth]{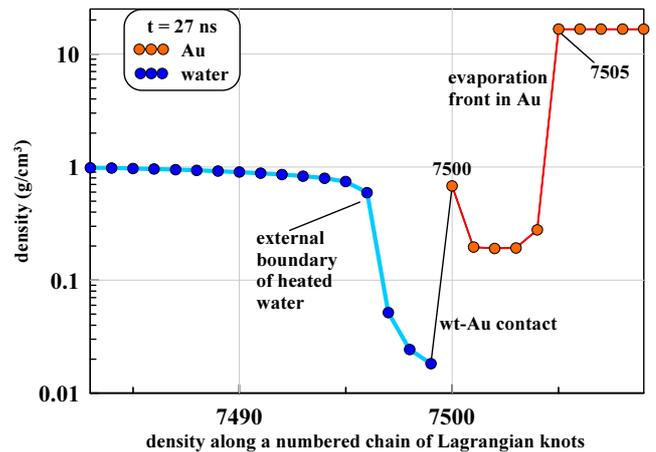}
\caption{Mass distribution in heated layers of gold and water.
The numbering of Lagrangian knots goes from water to gold.
Therefore, in this figure, the gold is on the right, compared with the previous figures.
The parameters of the laser pulse are shown in Figure 1.
}
\label{fig:5}       
\end{figure}


 Let us estimate the mass of evaporated gold and highly heated water.
 The calculation is carried out by the code 1D-2T-HD that works in Lagrangian coordinates.
 It is necessary to cover the extended calculation region with a sequence of Lagrangian knots.
 The extent of this region grows and grows as the shock wave propagates farther and farther into the water.
 To cope with this difficulty, it is necessary to roughen the Lagrangian grid over time.
 We keep the number of Lagrangian knots approximately equal to $10^4.$


 The enlargement of Lagrangian segments $\Delta x^0$ occurs several times during simulation; when we are counting up to great times.
 The enlargement or increase means that more mass enters the Lagrangian segment $\Delta x^0.$
 The length $\Delta x^0$ corresponds to the length of the segment on the axis normal to the target surface before the start of the laser pulse.
 At the beginning of the simulation, the step $\Delta x^0$ along the Lagrangian grid was 1 nm in both gold and water.
 This means that the column mass (mass on the line of sight) $\rho_{Au} \Delta x^0$ and $\rho_{wt} \Delta x^0$
   is concentrated in the Lagrangian segment $\Delta x^0$ for gold and for water, respectively;
    here $\rho_{Au}$ and $\rho_{wt}$ are densities of Au and water at room temperature.

\begin{figure}
  \centering \includegraphics[width=\columnwidth]{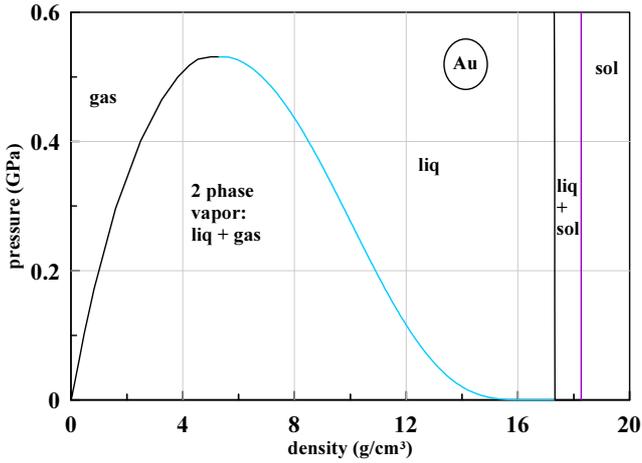}
\caption{Figure presents a phase diagram of gold on the plane density-pressure.
Phase boundaries are obtained in accordance with the wide-range equation of state
 \cite{Bushman:1993,Khishchenko2002,lomonosov_2007,rusbank1,rusbank2}.
Near the triple point, the saturation pressure drops well below 1 bar.
Therefore, the boiling curve presses exponentially strongly against the horizontal axis
 if we use a linear scale for pressure.
}
\label{fig:6}       
\end{figure}


 Figure \ref{fig:5} shows the density distribution along the Lagrangian coordinate at a time point of 27 ns;
   compare with figures \ref{fig:1} and \ref{fig:3}.
 The horizontal axis in this figure represents the numbers of the Lagrangian nodes in the chain.
 Therefore the circles/nodes in figure \ref{fig:5} are equidistant.
 Lagrangian knot number 7500 is the last Lagrangian particle of gold on the border with water in figure \ref{fig:5}.
 At the time shown in the figure, the length of the Lagrangian segment $\Delta x^0$ in water is 10 nm.
 As stated, a mass of water $\rho_{wt} \Delta x^0$ is concentrated in one Lagrangian particle.
 In a heated layer of water there are four such Lagrangian cells, see figure \ref{fig:5}.
 Therefore, the mass of heated water per unit area of the laser spot is equal to $4\,\rho_{wt} \Delta x^0$ $=0.4 \cdot 10^{-5}$ g/cm$\!^2.$
 Corresponding mass of hot water is $0.8\cdot 10^{-8}$ g for laser beam diameter $2 R_L=0.5$ mm.
 This is the mass of $3\cdot 10^{14}$ water molecules.


 About 4.5 Lagrangian cells make up a layer of evaporated gold, see figure \ref{fig:5}.
 At the time point 27 ns in the figure \ref{fig:5}, the length $\Delta x^0$ of one Lagrangian segment of gold is 2 nm.
 Therefore, the mass of evaporated gold per unit area is $=1.7 \cdot 10^{-5}$ g/cm$\!^2.$
 This mass is approximately four times the mass of hot water.
 Mass of evaporated gold is $3.4\cdot 10^{-8}$ g for the 0.5 mm beam; corresponding number of Au atoms is $10^{14}.$
 If these atoms condense into nanoparticles (NP) with a diameter of 4 nm,
   then the number of produced NPs in one laser shot will be $5\cdot 10^{10}.$

\begin{figure}
  \centering \includegraphics[width=\columnwidth]{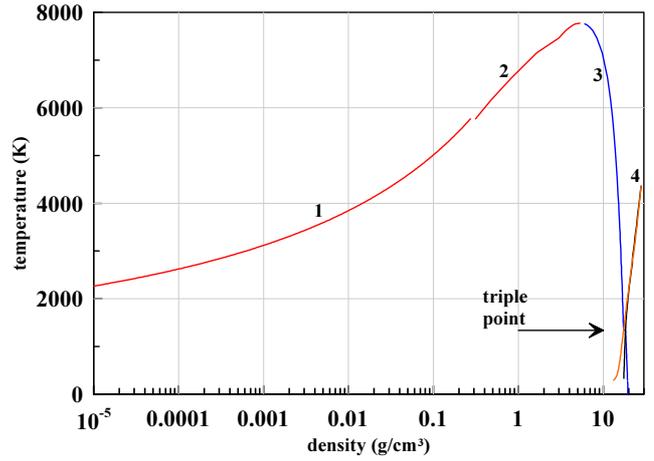}
\caption{The density-temperature phase plane of gold is shown.
Here 1 and 2 present the condensation curve,
 3 is the boiling curve, and 4 are solidus and liquidus bounding the two-phase liquid-solid mixture corridor,
  compare this corridor with the corridor "liq+sol" in figure \ref{fig:6}.
 The corridor in this figure looks narrow. This is due to the horizontal axis selected in this figure.
}
\label{fig:7}       
\end{figure}

\section{Parameters of evaporated gold}
\label{sec:3}


 Let's see how evaporation looks on the phase plane of gold.
 First, let's see how the phase planes are arranged.
 After that, we show the location of the hydrodynamic profile relative to the boundaries
   of the first-order phase transitions (see Section \ref{sec:4} below).

 Figure \ref{fig:6} shows binodal (coexistence curve) and two-phase melting region.
 Expression
 \begin{equation} \label{eq:01Psat-T}
 p_{sat}(T) = 35 \exp(- \frac{39950}{T} + \frac{T}{11000} + \frac{T^2}{2.45\cdot 10^8} )
 \end{equation}
 approximates dependence of saturation pressure on temperature which follows the wide-range equation of state (EoS)
 \cite{Bushman:1993,Khishchenko2002,lomonosov_2007,rusbank1,rusbank2}.
 Critical parameters for this version of EoS are:
 \begin{equation} \label{eq:01a-crit}
 \rho=5.3 \, {\rm g/cm}\!^3, \;\; T\approx 7800 \, {\rm K,} \;\; p=5300 \, {\rm bar.}
 \end{equation}
 The approximation of the boiling curve has the form:
\begin{eqnarray}
\label{eq:02rhoBOIL-T}
\nonumber 
 \rho(T) &=& 5.3 + 0.177 \sqrt{7774-T}-1.4\cdot 10^{-4}(7774-T)- \\
         & &       3.1\cdot 10^{-8}(7774-T)^2, \\
\nonumber 
 T(\rho) &=& 7800 - 39.2(\rho-5.3)^2 + 1.9(\rho-5.3)^3 - \\
         & &        0.195(\rho-5.3)^4,
\end{eqnarray}
where density $\rho$ is in g/cm$\!^3,$ temperature in K.

\begin{figure}
  \centering \includegraphics[width=\columnwidth]{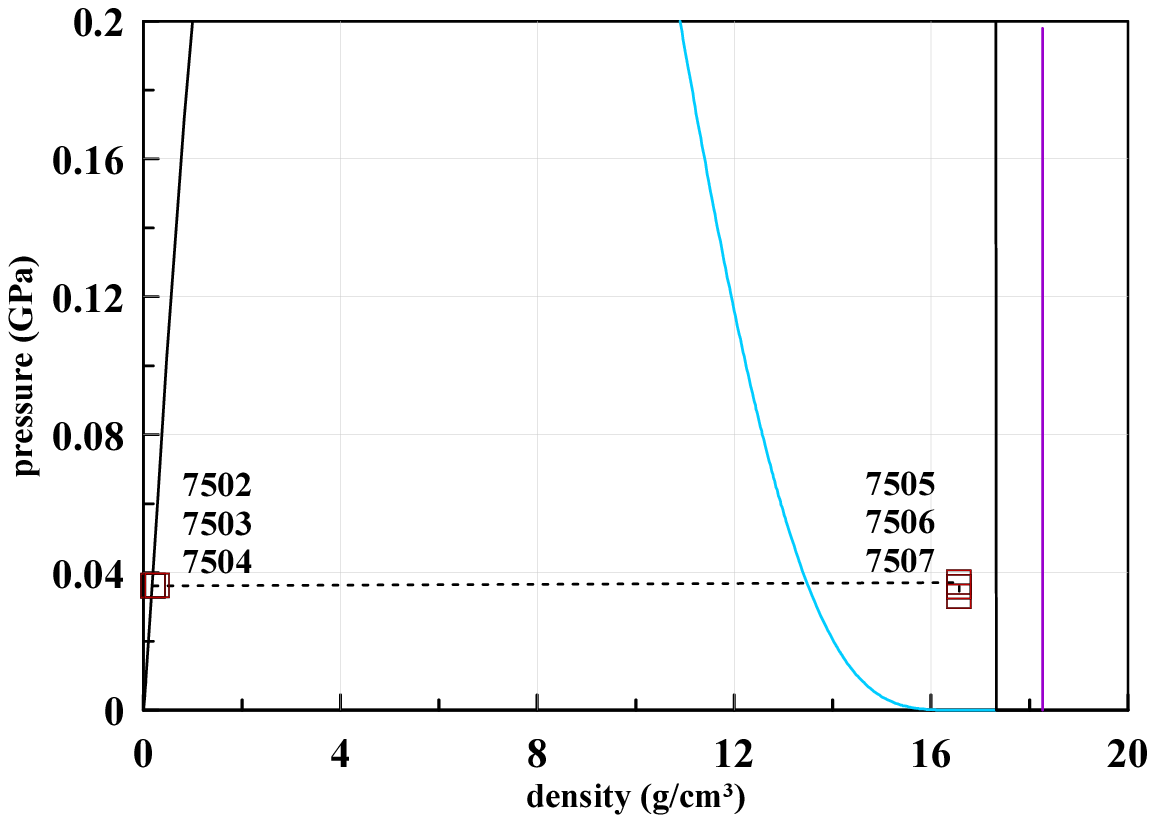}
\caption{The hydrodynamic profile of gold from figures \ref{fig:4} and \ref{fig:5}
 is plotted on the phase diagram from Figure \ref{fig:6}.
 The indicated numbers of Lagrangian knots 7502 ... correspond to Figure 5.
 We see what jump in density occurs at the evaporation front.
 This jump connects the region of liquid phase and the condensation curve.
 The water region, the gold vapor and the liquid phase region are under a pressure of 360 bar.
 Therefore, Lagrangian nodes number 7505 and higher are located in the depth of the liquid phase,  away from the boiling curve.
}
\label{fig:8}       
\end{figure}

\begin{figure}
  \centering \includegraphics[width=\columnwidth]{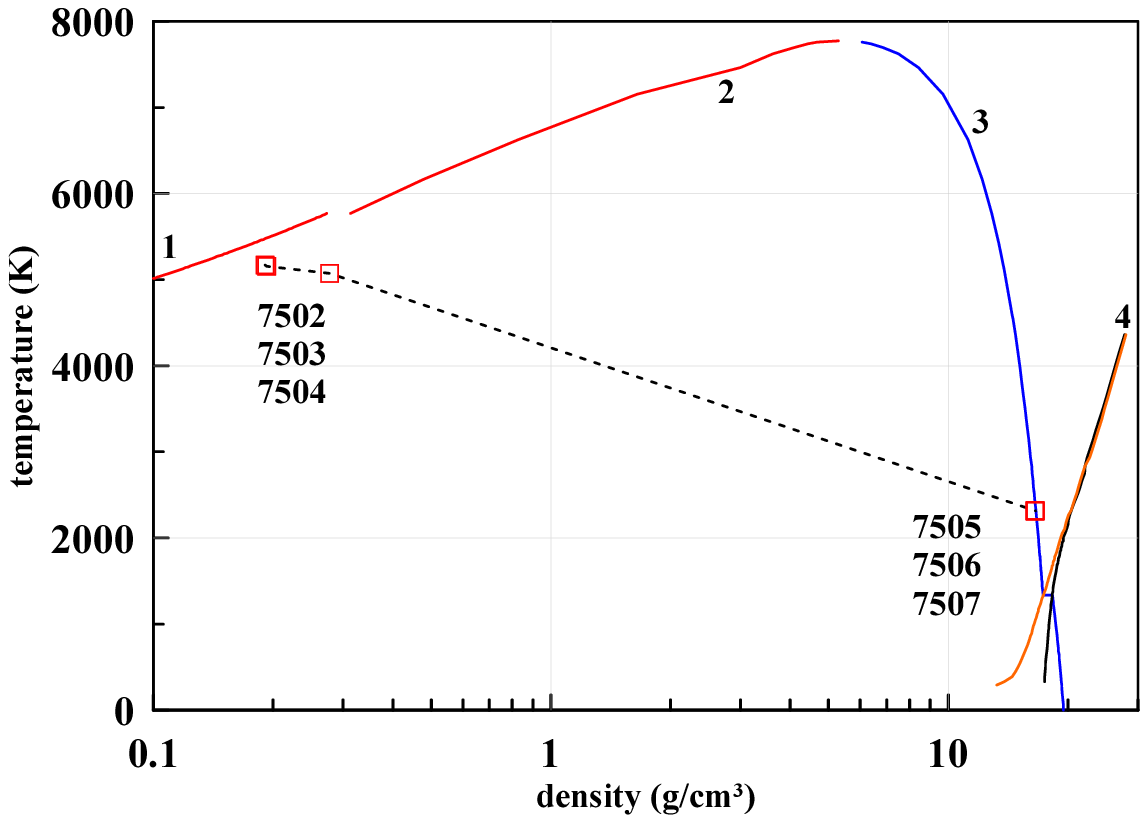}
\caption{The hydrodynamic profile of gold from figures \ref{fig:4} and \ref{fig:5}
 is plotted on the phase diagram from Figure \ref{fig:6}.
 The indicated numbers of Lagrangian knots 7502 ... correspond to Figure 5.
 We see what jump in density occurs at the evaporation front.
 This jump connects the region of liquid phase and the condensation curve.
 The water region, the gold vapor and the liquid phase region are under a pressure of 360 bar.
 Therefore, Lagrangian nodes number 7505 and higher are located in the depth of the liquid phase,
  away from the boiling curve.
}
\label{fig:9}       
\end{figure}


 Excluding temperature $T$ from the parametric dependencies (\ref{eq:01Psat-T}) and (\ref{eq:02rhoBOIL-T}),
  we obtain the right branch of the binodal (the boiling curve $p_{boil}(T))$ in the density - pressure plane.
 This branch is shown in figure \ref{fig:6} as the blue curve.
 The left branch in the figure \ref{fig:6} is drawn according to the wide-range EoS
 \cite{Bushman:1993,Khishchenko2002,lomonosov_2007,rusbank1,rusbank2}.
 This is the condensation curve, which is highlighted in black in the figure \ref{fig:6}.
 The asymptotic behavior at the zero $\rho=0, p=0$ of the condensation curve has the form
 $$
 \rho \propto p \ln (p_*/p), \;\;\; p\to 0, \;\; p \ll p_*.
 $$
 This means that the condensation curve $p_{cond}(\rho)$ on the plane $\rho,p$
    weakly (only logarithmically) touches the horizontal axis: $d p_{cond}(\rho)/d\rho \to 0$ as $\rho\to 0.$
 This touch is so weak that in the figure \ref{fig:6} the dependence $p_{cond}(\rho)$ near zero looks like a straight line.


 Figure \ref{fig:7} shows the situation on the plane of density and temperature.
 Curves 2, 3, and 4 in this figure are constructed according to the wide-range equation of state (EoS)
   \cite{Bushman:1993,Khishchenko2002,lomonosov_2007,rusbank1,rusbank2}.
 The tabular wide-range equation of state goes in rather large steps in the region of a substantially rarefied gas.
 In the figure \ref{fig:7}, the coexistence curve of gas and a two-phase mixture is constructed
  using formulas for saturated vapor pressure (\ref{eq:01Psat-T}) and equations of state of an ideal gas:
   $\rho_{cond}(T) =(M/k_B)p_{sat}/T;$ the curve 1; here $M$ is mass of gold atom.
 This is done in order to increase the accuracy of the representation of the condensation curve.

 The compressibility factor $p_{ideal}/p_{cr}$ for wide-range EoS is 3.3, where $p_{ideal}=(\rho_{cr}/M) k_B T_{cr};$
  critical parameters are presented in (\ref{eq:01a-crit}); they are marked with a subscript "cr" here.
 This factor characterizes, firstly, the amplitude of the cohesive forces at the critical point
  and, secondly, the error of the equation of state of the ideal gas at the critical point.
 The specified error decreases with distance from the critical point.
 Therefore, curve 1 in Figure \ref{fig:7} is consistent with the tabular wide-range EoS
   as the temperature decreases from the critical temperature (\ref{eq:01a-crit}) to a temperature of $\approx 6000$ K.

\section{Thermodynamic characteristics of the hydrodynamic profile in the evaporated layer}
\label{sec:4}



 The thermodynamic description of gold was presented above (Section \ref{sec:3}).
 Let us see in what thermodynamic states the portions of gold are in the case of the hydrodynamic profile
  shown in the figures \ref{fig:1}-\ref{fig:5}.
 The combination of the profile and the phase diagram is shown in figures \ref{fig:8} and \ref{fig:9}.
 The profile is represented by its Lagrangian nodes 7502-7507.
 The nodes with these numbers refer to evaporated gold and to the first few Lagrangian particles of gold melt,
   see figure \ref{fig:5}.


 From the analysis of figures \ref{fig:8} and \ref{fig:9} it follows that the distribution of quantities
   along the hydrodynamic profile strictly corresponds to the thermodynamic characteristics of substances.
 It is important that the method we used, together with the tabular equations of state,
  allows us to cross-cut through the fronts of first-order phase transitions.
 Through, i.e. without highlighting the front and dividing the flow into two parts -
  on one and the other side of the front.
 In this case, the coordination of heat fluxes to the left and to the right of the front
  and the front displacement velocity occurs automatically.
 Such coordination is necessary in order to provide the required heat flux
  to overcome the latent heat of fusion or evaporation.



 At the evaporation front, there are jumps in density and temperature, see figures \ref{fig:8} and \ref{fig:9}.
 The temperature jump is associated with a sharp drop in the thermal conductivity of gold
  during the transition of gold to a low-density phase.
 At the moment of 27 ns, the density in the vapor layer of gold decreases by two orders of magnitude
  compared to the value of 19.3 g/cm$\!^3.$
 Reducing of thermal conductivity plays an important role.
 As a result, the conductive cooling of gold steam slows sharply.
 The main cooling mechanism is adiabatic expansion.



 Basically, the expansion of the heated layers is supported by the pressure of saturated gold vapor
   together with pressure of hot water.
 At time 27 ns, shown in figures \ref{fig:8} and \ref{fig:9}, the pressure in the vicinity of the contact is 360 bar.
 There are the contact layers of cold and hot water and evaporated and liquid gold under this pressure,
  see figures \ref{fig:2} and \ref{fig:3}.



 In figure \ref{fig:8}, the edge of liquid gold (Lagrangian particles number 7505 and higher)
  is raised to a substantial height above the boiling curve (the blue curve).
 Whereas in figure \ref{fig:9} this edge is located directly on the boiling curve.
 The fact is that the pressure of 360 bar is very small compared to the bulk modulus of gold; 180 GPa in solid state.
 Accordingly, a change in density in the direction of compression is negligible and on the plane $\rho, T$
  the points related to the liquid, are right on the boiling curve, see figure \ref{fig:9}.
 The pressure rises sharply when we move from the boiling curve to a region
  occupied by liquid in the phase plane in figure \ref{fig:9}.

\begin{figure}
  \centering \includegraphics[width=\columnwidth]{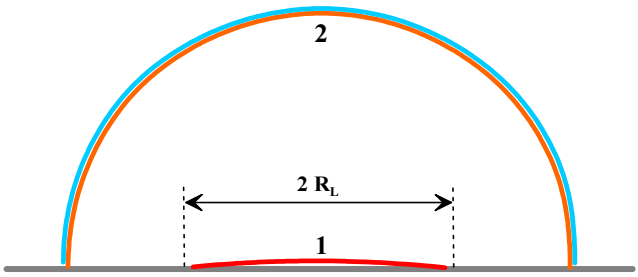}
\caption{Transformation of a plane hot layer 1 near a contact into a hemispherical bubble 2.
Transformation occurs at times of the order of 0.1-1 microsecond.
The boundary 2 of bubble separates hot water gas inside a bubble and cold liquid water surrounding the bubble.
The bubble expands until it reaches its maximum radius $R_{max}(t_{max})$ at a time point of $t_{max}.$
Diameter of an illuminated spot is $2R_L.$}
\label{fig:10}       
\end{figure}


 As was said, in figure \ref{fig:8}, the edge of liquid gold is raised to a substantial height above the boiling curve.
 This is because there is very little saturated vapor pressure related to density 16.6 g/cm$\!^3$
   and temperature $T=2313$ K at the edge of the liquid; $p_{sat}(T=2313)=14$ millibar $\ll 360$ bar.
 This means that the vaporized gold layer is formed at much earlier and much hotter stages of evolution
   compared with the time moment of 27 ns presented in figures \ref{fig:8} and \ref{fig:9}.
 The mass additions to the vaporized gold layer can be neglected at stages of the order of ten nanoseconds
  because evaporation rate drops drastically.

\begin{figure}
  \centering \includegraphics[width=\columnwidth]{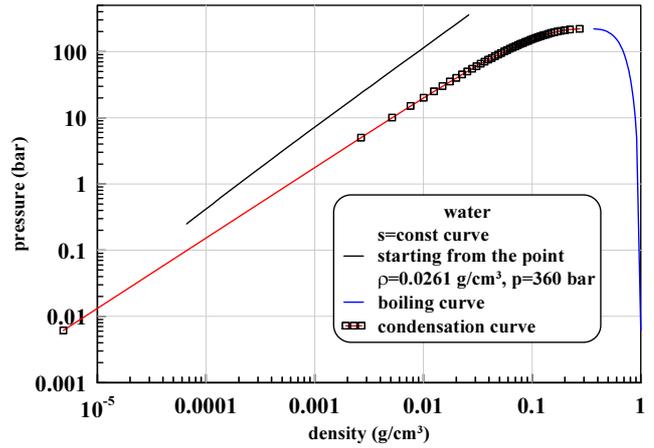}
\caption{ This figure presents the binodal of water and the adiabat of water.
 The adiabat begins in time 27 ns in the point with a pressure of 360 bar and a density of 0.0261.
 The stating point corresponds to water in a heated layer at a time 27 ns, see figures \ref{fig:4} and \ref{fig:5}.
 The adiabatic curve shows evolution of the heated layer down to the stage when pressure in hot water drops down
  to values below the ambient pressure 1 bar in the experimental container.
 During this evolution the hot layer transforms from plane to semi-spherical geometry as it is shown in figure \ref{fig:10}.
 The semi-spherical bubble expands up to the stopping stage where its expansion changes to contraction.
 Near this stage the lowest pressures inside the bubble are achieved.
 }
\label{fig:11}       
\end{figure}

\section{Role of water. Adiabatic expansion of water}
\label{sec:5}



At times of the order of 100-1000 ns, a hemispherical bubble 2 gradually forms
    from a quasi-flat hot layers of vaporized gold and gaseous water shown in figures \ref{fig:1}-\ref{fig:5}.
  This process is depicted in figure \ref{fig:10}.
  Hot layers 1 in figure \ref{fig:10} are near the contact.



 The name "cavitation" bubble is widely used in the literature on LAL.
 In fact, the bubble does not form at all due to cavitation.
 It is directly inflated with evaporated water.
 In the early stages, water heated by contact with hot metal evaporates.
 Further, cold water recedes under the pressure of hot gaseous water.
 That is, the bubble is correctly called the boiling bubble.



 Moreover, the above applies to laser ablation in a fluid (LAL) using femtosecond (fs) laser action (fs LAL).
 In this case, in a early short time interval, there is indeed cavitation inside the target material.
 But still, inflation of the bubble occurs much later.
 This late stage expansion of the bubble in liquid is again due to the expansion of gaseous ablation products.



 Figure \ref{fig:11} shows the binodal of water and the adiabat of water.
 This adiabat starts from a point with a pressure of 360 bar and a density of 0.0261 g/cm$\!^3.$
 The indicated point corresponds to the state of water in a highly heated layer
   at a time point of 27 ns, see figures \ref{fig:4} and \ref{fig:5}.
 The boiling and condensation curves in Figure \ref{fig:11} are taken from NIST reference data \cite{NIST}.
 The adiabatic curve starting from the point $\rho=0.0261$ g/cm$\!^3,$ $p=360$ bar was calculated
   using water steam calculator \cite{watersteamcalculator}.





 Below we estimate the maximum radius $R_{max}$ of expansion of the bubble, see figure \ref{fig:10}.
 At this value of the radius, the expansion of the bubble stops.
 Further, the bubble is compressed by a low external pressure.
 External pressure is small compared to the pressure in the contact layer in the early stages.
 Let the external pressure $p_{ext}(t_{max})$ at the moment $t_{max}$ of maximum expansion of the bubble
    equals to the pressure $p_0$ in the experimental vessel before the laser exposure: $p_{ext}(t_{max})=p_0.$
 Let the pressure $p_0$ in this vessel be 1 bar prior to exposure.


 From the solutions of the Rayleigh-Plesset equation,
   it is known that at the moment $t_{max}$ of inhibition of motion around the bubble
    the pressure inside the bubble $p_{in}(t_{max})$ drops below external pressure $p_{ext}(t_{max}).$
 This is due to the inertia of the expanding fluid that surrounds the bubble.
 In this case, the internal pressure $p_{in}(t_{max})$ in the bubble
   is several times lower than the external pressure $p_{ext}(t_{max});$
    this is overexpansion beyond the point of equilibrium of forces, as in the case of a pendulum.
 For these reasons, the end point $\rho_{in}(t_{max}), p_{in}(t_{max})$
   for calculating the expansion adiabat in Fig. \ref{fig:11} was chosen.



 The analysis of the question of the value of $p_{ext}(t_{max})$ requires a separate discussion,
   which is beyond the scope of this article.
 A simplified version $p_{ext}(t_{max})=p_0$ is adopted here.
 Let us describe at a qualitative level what the complexity is.
 A shock wave propagates around the surrounding fluid, see Figure \ref{fig:2}.
 There is pressure $p_0$ in front of this shock wave (pressure $p_{SW}$ is just behind the front)
   and pressure $p_{ext}(t)$ in the depth behind the wave in the region of the contact layer.
 Pressure $p_{ext}(t)$ is internal for the shock wave and external
   first for the hot plane layer and then for the bubble.
 The pressure in bubble $p_{in}(t)$ is different from pressure $p_{ext}(t)$ at distances of several radii of the bubble.



 The pressure $p_{ext}(t)$ behind the shock wave can drop significantly below the pressure $p_0$
   in front of the shock front;
    this is formation of a rarefaction cavity behind a strong shock wave.
 If this occurs at the stage of maximum expansion of the bubble,
  then pressure $p_{ext}(t_{max})$ will be significantly lower than pressure $p_0;$
   because the bubble is inside the rarefaction cavity in this case.
 In this case, it will be necessary to extend the adiabatic curve in Figure \ref{fig:11} further,
  in the direction of increased extensions.



 What the situation is with function $p_{ext}(t)$ depends on the size of the vessel $L_{ves}$ and bubble $R_{max};$
   of course, we consider the case $R_{max}\ll L_{ves}.$
 If the vessel is small, then there is a reflection of the shock wave from the walls of the vessel,
  scattering of the reflected waves,
   and pressure drop in the vicinity of the bubble $p_{ext}(t)$ to $p_0;$
    weakly non-linear shock propagates in water with speed of sound 1.5 km/s and passes 15 cm during 100 $\mu$s.
 But if (a) the vessel is large $L_{ves}\gg R_{SW}(t_{max})$
  and (b) the excess pressure $p_{SW}-p_0$ in the shock wave does not become small
   $p_{SW}-p_0 \gg p_0$ at the stage of maximum expansion $t\sim t_{max}$ of the bubble,
    then pressure $p_{ext}(t)$ will drop below pressure $p_0.$



 We conclude this discussion of the question of pressure $p_{ext}(t_{max}).$
 We restrict ourselves to the approximation $p_{ext}(t_{max})=p_0.$
 This approximation with $p_0=1$ bar corresponds to the endpoint of the adiabatic curve in figure \ref{fig:11}.

\section{Expansion of a bubble}
\label{sec:6}


 Determine the size of $R_{max}=R(t_{max})$ from the expansion degree along the adiabatic curve;
   here $R(t)$ is radius of a bubble.
 We write the law of conservation of mass:
 \begin{equation} \label{eq:04Rmax}
 \pi R_L^2 d_{hot} \rho_{hot} = (1/2)(4/3)\pi R_{max}^3 \rho_{in}(t_{max}),
 \end{equation}
  here $R_L$ is defined in figure \ref{fig:10},
   $d_{hot}$ is thickness of the hot contact layer shown in figures \ref{fig:1} and \ref{fig:4},
   $\rho_{hot}$ is density in the point where the adiabatic curve in figure \ref{fig:11} starts,
   and density inside a bubble at the stage of maximum expansion $\rho_{in}(t_{max})$
    is density in the point where the adiabatic curve in figure \ref{fig:11} ends.
 The law (\ref{eq:04Rmax}) holds, since entropy changes steeply from the hot layer to cold water,
  see figures \ref{fig:1}, \ref{fig:4}, and \ref{fig:5}.
 Therefore, the evaporation/condensation exchange between the gas in the bubble and the surrounding liquid is small.



 From mass conservation (\ref{eq:04Rmax}) follows the formula
 \begin{equation} \label{eq:05Rmax}
 R_{max} = \{(3/2) \, [\, \rho_{hot}/\rho_{in}(t_{max})\,]\, d_{hot}\}^{1/3} R_L^{2/3}.
 \end{equation}
 We take the values: $\rho_{hot}=0.03$ g/cm$\!^3,$
  $\rho_{in}(t_{max})= 7\cdot 10^{-5}$ g/cm$\!^3$ (this is the end point in figure \ref{fig:11}),
  $d_{hot} = 2$ microns (this is thickness of the hot layer from gold and water in figure \ref{fig:4}).
 In this case, the equation (\ref{eq:05Rmax}) takes the form:
 \begin{equation} \label{eq:06Rmax}
 R_{max}\, [{\rm mm}] = 0.23\, (R_L/100 \, \mu {\rm m})^{2/3}.
 \end{equation}




 For the production of colloids of nanoparticles, rather large spots $2R_L$ (for illustration see figure \ref{fig:10})
  of illumination of the order of a millimeter are used.
 The size of $R_{max}$ is usually on the order of a millimeter.
 Of course, the size of a bubble substantially depends on the size of the spot.
 Large bubbles are easier to detect.
 The estimate (\ref{eq:06Rmax}) gives $R_{max}$ significantly less than 1 mm.
 Apparently, this is due to the relatively small absorbed energy $F_{abs}=0.9$ J/cm$\!^2,$ adopted in our calculation;
   calculation parameters are shown in figure \ref{fig:1}.
 In addition, in experiments the pulse duration $\tau_L$ usually is longer, than our, often 5-10 ns.
 Then, firstly, the hot layer in figure \ref{fig:4} will become even hotter, and secondly,
  the hot layer in figures \ref{fig:1}, \ref{fig:4}, and \ref{fig:5} will become thicker.



 Even in our not extremely hot case the whole adiabatic curve corresponding to a hot layer
   passes above the two-phase region of water during all stages of evolution, see figure \ref{fig:11}.
 It is sometimes believed that water immediately enters the two-phase region
   as soon as the pressure drops below a critical water pressure of 220 bar.
 In our case, this is not so, see figure \ref{fig:11}.
 It also means that the effective adiabatic exponent
 \begin{equation} \label{eq:07gamma}
 \gamma = d \ln p|_S/d \ln \rho
 \end{equation}
  of expanding gaseous water is noticeably greater than unity;
    here $p|_S(\rho)$ is the function presented as the adiabatic curve in figure \ref{fig:11}.
 With good accuracy we have $\gamma \approx 1.24$ (\ref{eq:07gamma}) for the adiabatic curve shown in figure \ref{fig:11}.
 Whereas inside the two-phase region at low mixture densities, the pressure on the adiabatic curve
  changes with an exponent $\gamma$ that is only slightly greater than unity.



 The starting point of the adiabatic curve in figure \ref{fig:11} corresponds to a time of 27 ns
   in a particular calculation, presented above (figures \ref{fig:1}-\ref{fig:5}, \ref{fig:8}, \ref{fig:9}, \ref{fig:11});
     calculation parameters are shown in figure \ref{fig:1}.
 Initially, a decrease in density and pressure in the hot layer occurs due to planar expansion,
   that is, due to the expansion in direction normal to the surface of the target.
 Then there is a transition to spherical expansion, see figure \ref{fig:10}.
 At what point in time $t_{sph}$ the spherization of expansion of the bubble occurs depends on the radius $R_L.$
 The larger the radius, the later such a transition will take place.



\begin{figure}
  \centering \includegraphics[width=\columnwidth]{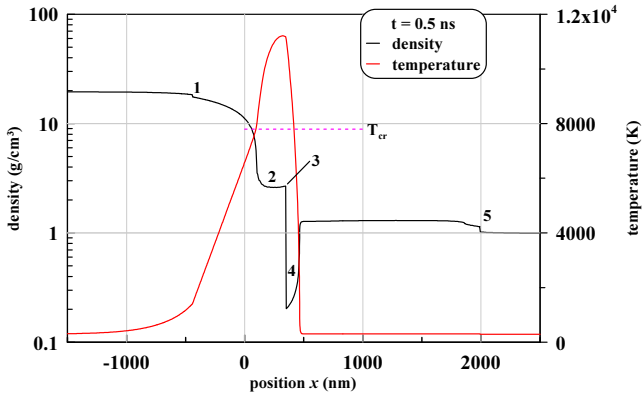}
\caption{The shape of the density and temperature profiles obtained thanks to hydrodynamic simulation 1D-2T-HD
 is shown.
 These profiles relate to the stage of completion of gold heating by a laser pulse.
 The pulse parameters are shown in figure \ref{fig:1}.
 Here, marker 1 shows the instantaneous position of the melting front.
 To the right of the front 1 is a molten gold.
 The following markers refer to a gold layer in a non-ideal gas state (marker 2),
 hot water layer (marker 4),
 the contact between gold and water is marked with the number 3,
 5 is shock wave in water.
 }
\label{fig:12}       
\end{figure}

\begin{figure}
  \centering \includegraphics[width=\columnwidth]{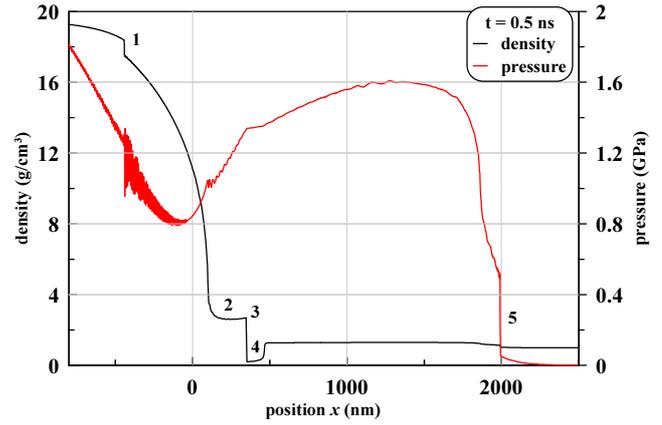}
\caption{The shape of the density and pressure profiles.
 These profiles relate to the stage of completion of gold heating by a laser pulse, see also figure \ref{fig:12}.
 Markers are: 1 is the melting front inside bulk gold,
 2 is hot rarefied gold,
 4 is hot water layer,
 3 is contact,
 5 is shock wave in water.
 Very noticeable is the minimum on the pressure profile.
 The appearance of a minimum is caused by the end of heating of gold with a laser pulse, see text for explanation.
 }
\label{fig:13}       
\end{figure}

\section{Early hydrodynamic evolution}
\label{sec:7}



 The results related to the time of 27 ns were presented above,
   figures \ref{fig:1}-\ref{fig:5}, \ref{fig:8}, \ref{fig:9}.
 At this point in time, the shock wave in the water travels to a distance of about 60 microns from the contact,
   see figure \ref{fig:2}.
 That is, at spot radii $R_L$ of the order of 100 microns,
   the stage of transition to the regime of spherical propagation of a shock wave in water approaches.
 At this stage, the one-dimensional (1D) approximation adopted in the code 1D-2T-HD becomes inapplicable.



 Here we are talking about the stage of spherical propagation of a shock wave;
   everywhere refers to the hemisphere.
 The transition to the stage of spherical expansion of the bubble occurs much later,
   since the speed of expansion of the bubble is much lower than the speed of sound in water (1.5 km/s).



 To describe the subsequent evolution, we use the adiabatic approximation, see figure \ref{fig:11}.
 The fact is that the thermal diffusivity of water in the hot layer becomes small,
   of the order of $\chi_{wt.hot} \approx (3-4)\cdot 10^{-3}$ according to reference data \cite{NIST}.
 In such a situation, the thermal cooling time $t_{wt.hot}$ of the hot water layer
 $$
 t_{wt.hot} \sim (1/4) \, d^2/\chi_{wt.hot} \sim 3\cdot 10^{-5}\, [{\rm s}]
 $$
  becomes about 3 microseconds; here we use thickness of the hot layer $d$ equal to 2 microns.
 Whereas the hydrodynamic expansion time $t_{hd}=d/u_{hd}$ of the hot layer at the stage
   shown in the figures \ref{fig:1}-\ref{fig:5} is about 0.1 microseconds.
 Indeed, at a point in time of 27 ns, shown in figure \ref{fig:2},
   the velocity of expansion of the hot layer is about 30 m/s.
 Let's also mention that in this time, the water velocity immediately behind the shock wave front is 180 m/s.

\begin{figure}
  \centering \includegraphics[width=\columnwidth]{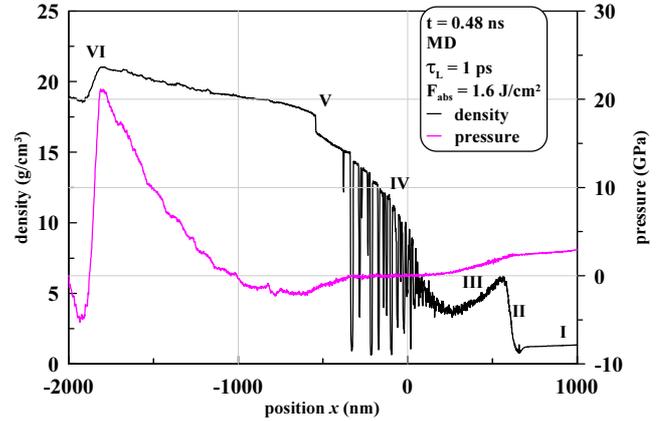}
\caption{The shape of the density and pressure profiles obtained in MD (molecular dynamics) simulation.
Parameters are: $F_{abs}=1.6$ J/cm$\!^2,$ $\tau_L=1$ ps.
Here I is shocked water,
II is the edge of an "atmosphere" and heated layer of diffusively mixed gold and water,
III is the "atmosphere",
IV is gold foam, pressure drops to zero in this foam layer,
V is melting front,
VI is rarefaction wave.
See text for explanations.
 }
\label{fig:14}       
\end{figure}

\begin{figure}
  \centering \includegraphics[width=\columnwidth]{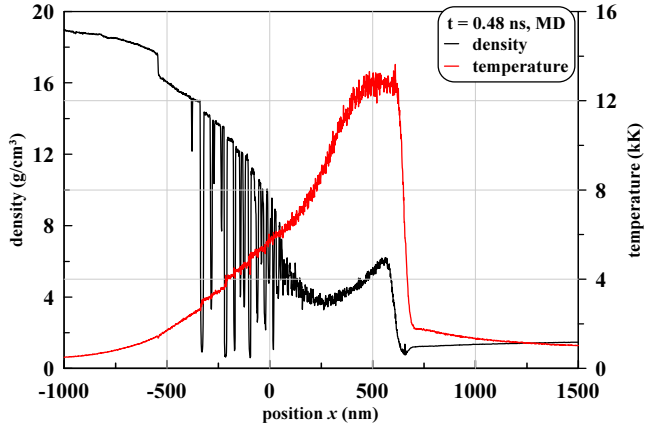}
\caption{The shape of the density and temperature profiles, compare with figure \ref{fig:14}
We see that temperature is high near the contact, compare with figure \ref{fig:12}.
 }
\label{fig:15}       
\end{figure}



 So, the data at the time point of 27 ns were presented above, figures \ref{fig:1}-\ref{fig:5}.
 This moment of time was used by us for the transition from the hydrodynamic code 1D-2T-HD
  to adiabatic evolution -
   the starting point on adiabatic curve in figure \ref{fig:11} relates to the state of hot water at time 27 ns.
 Let's give briefly here the data related to the initial stage of evolution.
 The pulse duration $\tau_L$ in the hydrodynamic simulation is 0.5 ns.
 The calculation parameters are shown in figure \ref{fig:1}.
 Figures \ref{fig:12} and \ref{fig:13} show the hydrodynamic profiles at a time moment of 0.5 ns,
   when the laser heating ends.



 A laser pulse passes through transparent water and is absorbed at the edge of a gold target.
 Further, thermal conductivity distributes the absorbed energy into the thickness of gold.
 At the same time, water is heated from gold in a contact way.
 The thermal conductivity of dense gold is greater than the thermal conductivity of water.
 Therefore, much more heat is stored in gold.
 The temperature distribution at time 0.5 ns is shown in Figure 12.



 Gold heating in the absorption zone is accompanied by a rise in pressure in this zone.
 Of course, under the influence of an ultrashort pulse,
  the pressure calculated for the same absorbed energy $F_{abs}$ is greater,
    than when exposed to a nanosecond pulse.
 However, the generation of significant pressures is also associated with a long pulse, see Figure \ref{fig:13}.
 An ultrashort pulse will be called a pulse for which the acoustic time scale $t_s=d_T/c_s$
  is shorter than the heating duration; here $d_T$ is thickness of heat affected zone, $c_s$ is speed of sound in gold.
 In the case of a long pulse, on the contrary, the acoustic scale $t_s$
  is shorter than the pulse duration $\tau_L.$
 For figures \ref{fig:12} and \ref{fig:13}, the thickness $d_T$ is approximately 700 nm.
 Accordingly, the acoustic scale $t_s$ is about 200 ps.



 The increased pressure leaves the heating zone at the speed of sound in the case of a long pulse.
 Two smooth pressure hillocks arise, the shape of which is associated with the profile of the laser pulse in time.
 In our case, we use a pulse whose shape in time is equal to the Gaussian function $I\propto \exp(-t^2/\tau_L^2),$
  in our 1D-2T-HD simulation $\tau_L=0.5$ ns, see figure \ref{fig:1}; this is e-folding time.
 These acoustic hillocks are emitted along their characteristics towards gold and towards water,
   see Figure \ref{fig:13}.
 Radiation ceases with the cessation of heating.
 Accordingly, the spatial width of the hillocks is $\sim c_s\, \tau_L.$
 This is how the minimum pressure is formed on the pressure profile in figure \ref{fig:13}.
 A low pressure zone is formed between pressure hills that propagate into gold and water.

\section{Comparison of hydrodynamic and molecular-dynamics simulations}
\label{sec:8}



 Molecular dynamics (MD) simulation results are shown in figures \ref{fig:14} and \ref{fig:15}.
 The molecular dynamics simulation technique is described in the article \cite{INA.jetp:2018.LAL}.
 In contrast to the hydrodynamic simulation described above (figures \ref{fig:1}-\ref{fig:5}, \ref{fig:12}, \ref{fig:13}),
   the case of ultrashort exposure with $\tau_L=1$ ps is considered here.
 Absorbed energy is $F_{abs}=1.6$ J/cm$\!^2.$

\begin{figure}
  \centering \includegraphics[width=\columnwidth]{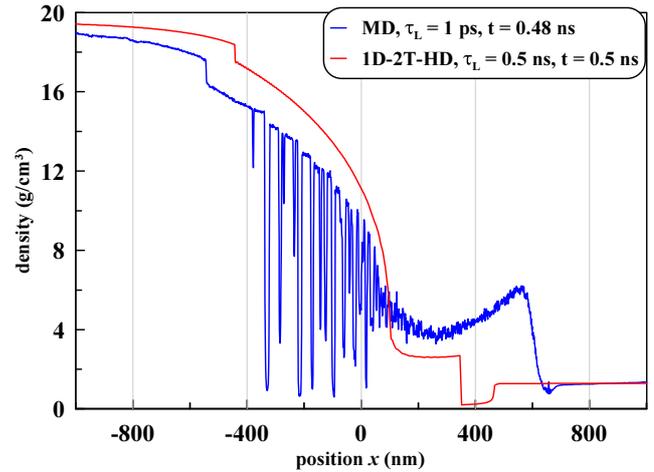}
\caption{Comparison of the 1D-2T-HD and MD profiles.
 }
\label{fig:16}       
\end{figure}



 Density and pressure profiles are shown in figure \ref{fig:14}.
 Shock in water at the right side isn't shown.
 The fact is that in molecular dynamics modeling we use a special boundary condition on the right side in the water.
 This condition is placed on the Lagrangian particle at a considerable distance from the contact.
 First, in the calculation with a small cross section, the trajectory of this Lagrangian particle is determined;
  the cross section is perpendicular to the axis $x$ along direction of motion.
 Then, in the MD calculation with a large cross section,
  a part of the water from the contact to the specified Lagrangian particle is used.
 This reduces the number of atoms that are used in the calculation.
 The technique is described in the article \cite{INA.jetp:2018.LAL}.
 Shock compressed water is marked by "I" in figure \ref{fig:14}.



 Due to nucleation under action of tensile stress at time $t\sim t_s$
   and due to separation of the spall layer III, a foam layer IV is formed, see figure \ref{fig:14}.
 In the initial stages, the spall layer moves at a considerable speed.
 When propagating into a liquid, this layer is slowed down thanks the inertia of the liquid,
   which is displaced by the spallation layer.
 This leads to a gradual slowdown in the movement of the spall layer.
 As a result, two circumstances are formed that are absent when the spallation layer expands into vacuum.
 Firstly, the spall layer is in quasi-hydrostatic equilibrium.
 Therefore, we call it the "atmosphere," see \cite{INA.jetp:2018.LAL} and figures \ref{fig:14}, \ref{fig:15}.
 The fact is that sound runs through this layer in a time that is less than the time during which braking occurs.
 This allows us to talk about hydrostatic equilibrium.
 As a result of the action of weight (effective gravity),
  pressure and density gradients in the atmosphere (hydrostatic compression) arise, see figure \ref{fig:14}.


\begin{figure}
  \centering \includegraphics[width=\columnwidth]{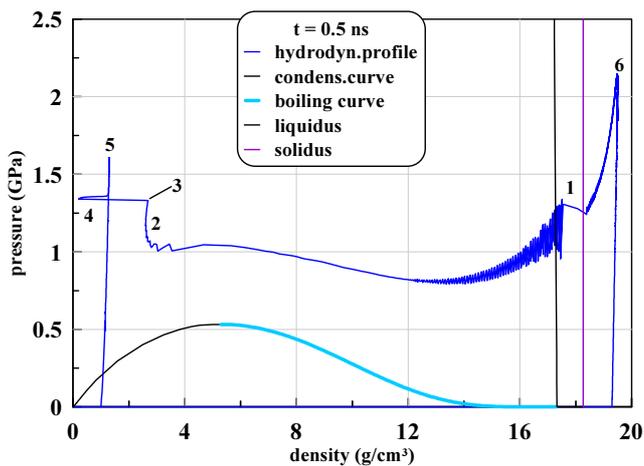}
\caption{This figure shows that at the initial stage,
 the substance in the hot layer 2 is located at the phase plane significantly higher
  than the critical point of gold.
 Under these conditions, the surface tension at the contact boundary with water disappears
  and the diffusion interpenetration of gold and water sharply intensifies.
 The numbers in the figure refer to the characteristic sections of the hydrodynamic profile,
   which are indicated by the same numbers in figures \ref{fig:12} and \ref{fig:13}.
 This is results of 1D-2T-HD simulation at time 0.5 ns.
 See the text for additional explanations.
 }
\label{fig:17}       
\end{figure}

\begin{figure}
  \centering \includegraphics[width=\columnwidth]{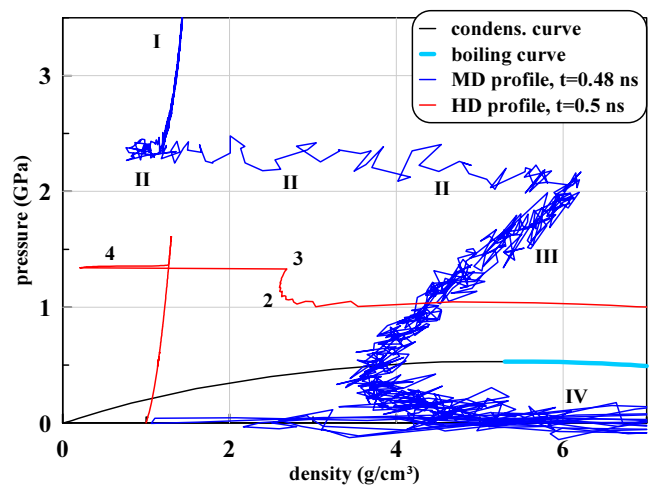}
\caption{This figure demonstrates that at the early stage,
 the substances in the hot layer 2 in HD
 and in layer II in MD are located at the supercritical region of the phase diagram.
 Thus the surface tension at the gold/water boundary is zero
  and there is effective diffusion interpenetration of gold and water.
 The numbers 4,3,2 in the figure refer to the characteristic sections of the HD profile,
   which are indicated by the same numbers in figures \ref{fig:12} and \ref{fig:13}.
 The Roman numerals I, II, III, IV refer to the characteristic sections of the MD profile
  shown in figure \ref{fig:14}.
 }
\label{fig:18}       
\end{figure}



 The second circumstance that distinguishes laser ablation in a liquid (LAL) is as follows.
 When flying into vacuum, the initial velocity distribution of the mass that has escaped is preserved over time.
 Therefore, the droplets formed during the decay of the foam fly behind fragments of the spall layer.
 Whereas when the spallation layer spreads into the liquid (LAL),
   the atmosphere is slowed down by the liquid.
 At the same time, the foam behind is protected from the liquid by the layer of the atmosphere and maintains its speed.
 As a result, membranes and droplets of foam fly into the atmosphere.
 There is an increase in mass concentrated in the atmosphere.
 In addition, the influx of momentum somewhat reduces the rate of deceleration of the atmosphere.
 In the end, most of the foam adheres to the atmosphere.



 Our gold layer used in the MD simulation has a finite thickness.
 On the left, this layer in figures \ref{fig:14} and \ref{fig:15} is bordered by vacuum (free boundary).
 At the time point 0.48 ns shown in figures \ref{fig:14} and \ref{fig:15},
  the shock wave in gold reflects from the free boundary.
 One can see how a rarefaction wave begins to propagate from the free boundary
   towards the triangular shock wave in gold.
 Dynamically, this is an insignificant detail if we are interested in the fate of the atmosphere,
  the diffusion layer, and shock-compressed water.



 Comparison of the 1D-2T-HD and MD profiles is presented in figure \ref{fig:16}.
 In 1D-2T-HD the hot layer of water is wider; maybe as a result of heat conduction in water included in 1D-2T-HD.
 Also comparing the figures \ref{fig:12} and \ref{fig:15},
  we see that in MD the shock wave leaves a more significant trace of heating in water.
 Of course, this is due to the fact that at the initial stage,
  shock waves created by ultrashort laser pulse are much more intense compared to nanosecond pulse.

\section{Comparison of hydrodynamic and molecular-dynamics thermodynamical states}
\label{sec:9}



 Figure \ref{fig:17} shows the thermodynamic states in the case of nanosecond exposure with parameters
  given in figure 1; simulation by the code 1D-2T-HD is presented.
 The time is 0.5 ns.
 The rise in pressure along the evaporated layer of gold 2
   is due to the formation of a minimum of pressure at the edge of the gold target.
 As mentioned above, the formation of a minimum is associated with the end of heating of gold by nanosecond laser pulse.
 As a result of this, the separation of acoustic waves traveling into gold and water occurs, see figure \ref{fig:13}.




 Numbers 1-5 indicate the characteristic sections of the hydrodynamic profile in figures \ref{fig:12} and \ref{fig:13}.
 These numbers are repeated in figure \ref{fig:17}.
 This is done in order to see which areas of the phase diagram belong to certain sections of the profile.
 The meaning of the number 5 in figures \ref{fig:13} and \ref{fig:17} is different.
 In figure \ref{fig:13}, this is the front of a forming shock wave in water.
 Whereas in figure \ref{fig:17} this is the maximum of the pressure hill in the water.
 A pressure hill is called a nonlinear acoustic wave traveling into water.

\begin{figure}
  \centering \includegraphics[width=\columnwidth]{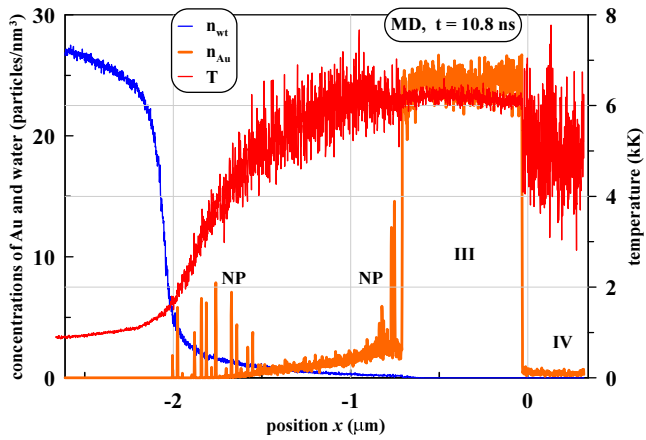}
\caption{Processes of diffusive mixing, cooling at the edge of the hot layer,
 and condensation of gold atoms into nanoparticles (NP) are shown here.
 }
\label{fig:19}       
\end{figure}


 In figure 17 the number 6 is added.
 The number 6 is not presented in figures \ref{fig:12} and \ref{fig:13}.
 The number 6 denotes a shock wave that propagates into the volume of a gold target.
 In figures \ref{fig:12} and \ref{fig:13} this wave is located beyond the left frame of the figures.
 As you can see, at the time 0.5 ns this wave runs on solid gold.



 Figure \ref{fig:18} is needed to demonstrate that in the simulations made in two different ways,
   the substance in the vicinity of the contact is in the supercritical region.
 Therefore, there is no surface tension and diffusive mixing of the target substance and the liquid proceeds rapidly.
 The hydrodynamic calculation was performed using the Lagrangian scheme.
 In this scheme, diffusion is absent.
 The presence of diffusion weakly affects the dynamics of the system
   (but is extremely important for production of nanoparticles).
 Molecular dynamics approach takes into account diffusion to the full extent.


 The Roman numerals I, II, III, IV in figure \ref{fig:18} refer to
  water compressed by compression wave (I, see figure \ref{fig:14}),
   to transition from water to gold (II),
   to "atmosphere" III,
   and to foam with negligible pressure inside (IV).


\section{Diffusion, condensation, and locking nanoparticles inside hot water}
\label{sec:10}




 Figure \ref{fig:19} shows, firstly, how the processes of diffusive interpenetration of gold and water proceed.
 Secondly, a temperature profile is presented.
 The thermal conductivity of rarefied water and low-density gold are very small.
 Therefore, a high temperature remains for a long time.
 The temperature decreases mainly due to the adiabatic expansion of the hot layer.


 Thirdly, we see the development of the process of condensation of atomic gold in a mixture with gaseous water.
 It is with this process that sharp spikes appear on the gold concentration profile
   $n_{Au}(x,t = 10.8 \, {rm ns}).$
 Groups of these spikes are marked with the letters "NP" (NanoParticles) in figure \ref{fig:19}.



 The pressure at time 10.8 ns shown in Figure \ref{fig:19} is approximately 700 bar.
 The Roman numerals I, II, III, IV in figure \ref{fig:18} refer to
 The numerals III and IV in figure \ref{fig:19} have the same meaning as in figures \ref{fig:14} and \ref{fig:18}.
 But figures \ref{fig:14} and \ref{fig:18} $(t=0.48$ ns) on the one hand
   and figure \ref{fig:19} $(t=10.8$ ns) on the other hand are far apart in time.
 Over the time interval from 0.48 ns to 10.8 ns, the area occupied by the foam was empty.
 Gold from foam IV sticks to the layer III.
 The layer of diffusion mixing expanded.



 The most important effect is as follows.
 Diffusing gold atoms fill a layer of low-density hot water.
 This is due to increased values of the diffusion coefficient in low-density water.
 At the edge of a layer of hot water, gold atoms begin to escape into cold water.
 The temperature of gold vapor in water decreases in this particular place.
 As a result, active condensation of gold atoms into nanoparticles begins.
 The spikes on the left edge of the diffusion layer in figure \ref{fig:19} correspond to it.



 So, on the edge that separates hot and cold water, the gold vapor condenses.
 Nanoparticles are involved in Brownian motion.
 Their mobility is small.
 The diffusion coefficient of nanoparticles is small compared with the diffusion coefficient of atoms;
   because they are much more massive and moves slowly.
 Consequently, the diffusion propagation of gold atoms into water is blocked
   at the interface between hot and cold water.



 The boundary between hot (high entropy) and cold water transforms into the boundary of the bubble
   when the water layer with high entropy expands.
 This process is shown in figures \ref{fig:10} and \ref{fig:11}.
 As a result, the gold nanoparticles are locked inside the bubble.
 This is precisely the situation observed in the experiments \cite{Kanitz_2019,LZ+Stephan:2018.LAL}.

\section{Conclusion}
\label{sec:11}

 The article is devoted to the analysis of processes that occur during laser ablation in a liquid.
 Two independent codes were used for analysis: the hydrodynamic code 1D-2T-HD and the molecular dynamics (MD) code.
 The complete evolution of the flow from the absorption of laser energy to the formation of a bubble in a liquid
   is considered. The main results are as follows.

 (A) At a sufficiently high absorbed energy of the order of Joules per square centimeter,
  the gold at the edge of the target appears for a rather long time in supercritical states,
   that is, in states above the critical point in temperature and pressure.
 Moreover, what has been said applies both to ultrashort pulses and to nanosecond pulses.

 (B) In supercritical states, capillary force disappears; this force separates substances
   at the contact boundary between them.

 (C) The disappearance of the dividing boundary sharply intensifies the diffusion interpenetration of substances.
The situation becomes similar to that which occurs when two gases come into contact.

 (D) Water near the border with gold is heated mainly due to thermal conductivity. There is also a contribution
related to the diffusion of hot gold vapors. Heating due to dissipation in the shock front is less significant.
Shock heating is greater when exposed to an ultrashort pulse, since the shock wave is much more powerful
compared to the nanosecond case. Thermal conductivity and diffusion create a thin layer of hot water in the
contact area.

 (E) The boundary between cold and hot water gradually transforms into the boundary of the bubble. Moreover, the
expanding hot water fills the inside of the bubble.

 (F) The diffusion of gold atoms is significant. Gold vapor fills a layer of hot water.

 (G) At the edge of hot and cold water, gold atoms condense into nanoparticles with sizes from a nanometer or more.
 The Brownian creep of the formed nanoparticles is slow.
 Their diffusion coefficient is small compared with the diffusion coefficient of atoms.
 Therefore, the nanoparticles are located inside the bubble.



\begin{acknowledgements}
The work of the INA, the VAKh, and YuVP was carried out under the state order of ITP RAS. VVZ thanks for support of the Russian Science Foundation grant number 19-19-00697.
\end{acknowledgements}


%
%

\bibliographystyle{spphys}       
\bibliography{flamn}   
\end{document}